\documentclass[twocolumn,english,pra]{revtex4-2}
\usepackage[T1]{fontenc}
\usepackage[latin9]{inputenc}
\usepackage{color}
\usepackage{babel}
\usepackage{amsmath}
\usepackage{amssymb}
\usepackage{esint}
\usepackage{graphicx}
\usepackage{hyperref}
\hypersetup{ 
           breaklinks=true,   
           colorlinks=true,   
           pdfusetitle=true,  
}

\makeatletter

\usepackage{braket}
\usepackage{babel}

\begin{document}
\title{Emergence of the molecular geometric phase from exact electron-nuclear
dynamics}
\author{Rocco Martinazzo$^{1,2,*}$, Irene Burghardt$^{3}$}
\affiliation{$^{1}$Department of Chemistry, Università degli Studi di Milano,
Via Golgi 19, 20133 Milano, Italy}
\email{rocco.martinazzo@unimi.it}

\affiliation{$^{2}$Istituto di Scienze e Tecnologie Molecolari, CNR, via Golgi
19, 20133 Milano, Italy}
\affiliation{$^{3}$Institute of Physical and Theoretical Chemistry, Goethe University
Frankfurt, Max-von-Laue-Str. 7, D-60438 Frankfurt/Main, Germany}
\begin{abstract}
Geometric phases play a crucial role in diverse fields. In chemistry
they appear when a reaction path encircles an intersection between
adiabatic potential energy surfaces and the molecular wavefunction
experiences quantum-mechanical interference effects. This intriguing
effect, closely resembling the magnetic Aharonov-Bohm effect, crucially
relies on the adiabatic description of the dynamics, and it is uncertain
whether and how it persists in an exact quantum dynamical framework.
Recent works have shown that the geometric phase is an artifact of
the adiabatic approximation, thereby challenging the perceived utility
of the geometric phase concept in molecules. Here, we investigate
this issue in an exact dynamical framework. We introduce instantaneous,
\emph{gauge} invariant phases separately for the electrons and for
the nuclei, and use them to monitor the phase difference between the
trailing edges of a wavepacket encircling a conical intersection.
In this way we unambiguosly assess the role of the geometric phase
in the interference process and shed light on its persistence in molecular
systems. 
\end{abstract}
\maketitle
\textbf{\emph{Introduction}}. Geometric phases represent fundamental
concepts in the realms of physics and chemistry. They are closely
associated with various phenomena, such as the quantum, the anomalous
and the spin Hall effect \citep{Girvin2019,VanderbiltBook2018}, the
exotic physics of topological insulators \citep{Hasan2010,Kane2013},
dielectric polarization in crystals \citep{Resta1992b,King-Smith1993b,Resta1994b,Resta2000,Xiao2010,VanderbiltBook2018},
the Aharonov-Bohm effect \citep{Aharonov1959}, and conical intersections
(CIs) in molecules \citep{Mead1992,Yarkony1996,Kendrick2003}. Typically
arising when the Hamiltonian of a system undergoes adiabatic changes
with respect to a set of ``environmental'' parameters $\mathbf{x}$
\citep{Berry1984}, geometric phases remain well defined even in scenarios
involving non-adiabatic, non-cyclic and non-unitary evolutions \citep{Simon1983,Aharonov1987,Simon1993,Mukunda1993}.
In the case of molecules, geometric phases play a critical role around
CIs between two or more potential energy surfaces. Their influence
is subtle and eludes a description in the commonly adopted  framework
defined by the Born-Oppenheimer approximation. This is the case even
if the molecular dynamics remains nearly adiabatic and does not occur
in the proximity of the CI. Geometric phases affect the quantum interference
of wavepackets encircling the CI \citep{Althorpe2006,Valahu2023},
no matter how close to the CI they move, and may have a significant
impact on the reaction dynamics \citep{Yuan2020,Kendrick2015}, although
cancellation effects may occur in the observables accessible in scattering
experiments \citep{Juanes-Marcos2005}. In these molecular problems,
the Berry phase is often not just geometric but also topological,
i.e., independent of both the dynamics and (to a large extent) of
the path. In fact, it is the phase introduced in the pioneering works
by Longuet-Higgins \citep{Longuet-Higgins1958} and by Herzberg and
Longuet-Higgins \citep{Herzberg1963}, and which is known to control
the ordering of the energy levels in Jahn-Teller systems. However,
these remarkable properties hinge crucially on the adiabatic approximation
of the dynamics, raising the question whether and how they persist
in an exact framework \citep{Min2014,Requist2016a}. 

Recent works \citep{Ibele2023,Martinazzo2023a} addressed this question
from a dynamical perspective, employing the same exact framework used
in Refs. \citep{Min2014,Requist2016a}, namely that provided by the
exact factorization (EF) of the wavefunction \citep{Abedi2010,Abedi2012}.
In our approach \citep{Martinazzo2023a} we exploited a \emph{gauge}-invariant
formulation of the EF dynamics recently derived from quantum hydrodynamics
(QHD) \citep{Martinazzo2023} to define geometric phases separately
for both the nuclei and the electrons, for arbitrary paths $\gamma$
in nuclear configuration space.  These phases, henceforth $\Gamma_{\text{n}}[\gamma]$
and $\Gamma_{\text{el}}[\gamma]$, respectively for nuclei and electrons,
always sum up to the phase difference of the \emph{total} electron-nuclear
($e-n$) wavefunction at the end points of the path. When the path
is closed the total phase difference vanishes \footnote{Strictly speaking this holds unless there is a node at the base point
of the path where the phase difference is considered. We shall address
this issue in more detail below.} and the phases are just the opposite of each other, i.e., $-\Gamma_{\text{n}}^{O}[\gamma]=\Gamma_{\text{el}}^{O}[\gamma]$
where the superscript ``$O$'' indicates that the path is a loop.
In this case, $\Gamma_{\text{el}}^{O}[\gamma]$ evolves in time according
to a (reversed) Maxwell-Faraday induction law, with non-conservative
forces arising from the electron dynamics that play the role of electromotive
forces \citep{Martinazzo2023a}. Furthermore, in the adiabatic limit,
these forces are conservative, $\Gamma_{\text{el}}^{O}[\gamma]$ becomes
stationary and the usual adiabatic geometric phase results. When the
path is not closed, however, $\Gamma_{\text{n}}[\gamma]$ and $\Gamma_{\text{el}}[\gamma]$
are \emph{distinct} contributions to the phase difference in the total
\emph{e-n} wavefunction, which can be assigned, respectively, to a
nuclear and an electronic \emph{wavefunction}. The \emph{gauge} choice
which is \emph{necessary} for this interpretation will emerge later
on, when analyzing these phases in detail.

The purpose of this work is to provide a time-dependent perspective
of the concept of geometric phase, i.e. to show how it emerges in
the course of an exact dynamics and how it affects, under suitable
conditions, the dynamics itself. To this end, we will focus on a model
Jahn-Teller problem and exploit $\Gamma_{\text{n}}[\gamma]$ and $\Gamma_{\text{el}}[\gamma]$
introduced above  to track the evolution of the phase differences
that develop between the trailing edges of a wavepacket encircling
the conical intersection. The path we choose is a ``dynamical''
path that moves in tandem with the nuclear probability density \textemdash{}
in accordance to a quantum hydrodynamical description of the exact
\emph{e-n} dynamics \citep{Martinazzo2023} \textemdash{} in such
a way that it is always ``at rest'' with the wavepacket itself and
its endpoints indeed follow the trailing edges of the latter. We show
that $\Gamma_{\text{n}}[\gamma]$ and $\Gamma_{\text{el}}[\gamma]$
entail key information about possible interference between different
portions of the wavefunction. This is close in spirit to the analysis
of geometric phase effects in adiabatic processes in which the slow
variables are (or can be treated as) external parameters that are
under full control of the experimenter. The key difference between
the latter and the present problem is that in the first case the quantum
vector describing the fast variables (the electrons) is the sole responsible
for the phase changes, while in the second the slow variables (the
nuclear degrees of freedom) participate in the dynamics and concur
in defining the quantum state under scrutiny.

\textbf{\emph{Non-adiabatic geometric phases}}. The exact factorization
of the molecular wavefunction \citep{Abedi2010,Abedi2012} extends
the fiber structure of the adiabatic approximation to arbitrary states,
thereby enabling a natural extension of the adiabatic Berry phase.
EF largely simplifies the description of the exact nuclear dynamics,
in fact very close to an adiabatic theory (a $U(1)$ \emph{gauge}
theory) if not for the presence of a time-dependent connection. By
contrast, the traditional Born-Huang expansion of the total wavefunction
\citep{Bohm2001} gives rise to a much more complicated theory \textemdash{}
for $n$ coupled electronic states, a $U(n)$ \emph{gauge} theory
\textemdash{} albeit with a stationary (but non-Abelian) connection
\citep{Wittig2012}. 

In the EF approach \citep{Abedi2010,Abedi2012} the wavefunction is
represented exactly as 
\begin{equation}
\ket{\Psi}=\int_{X}d\mathbf{x}\psi(\mathbf{x})\ket{u(\mathbf{x})}\ket{\mathbf{x}}\label{eq:EF wavefunction}
\end{equation}
where $\{\ket{\mathbf{x}}\}$ is the position basis of the nuclear
variables $x^{k}$ ($k=1,2,..N$), $\ket{u(\mathbf{x})}$ is the conditional
electronic state at $\mathbf{x}$ and $\psi(\mathbf{x})$ is the marginal
probability amplitude for the nuclei, i.e. the ``nuclear wavefunction''.
For each configuration $\mathbf{x}$ of the nuclei, the electronic
state and the nuclear wavefunction can be obtained from the amplitude
of the total $e-n$ wavefunction at $\mathbf{x}$, upon noticing that
the latter \emph{is} an electronic state, though generally not normalized.
Hence, upon imposing its normalization one defines the EF pair \{$\ket{u(\mathbf{x})}$,
$\psi(\mathbf{x})$\},
\begin{equation}
\braket{\mathbf{x}|\Psi}_{X}=\psi(\mathbf{x})\ket{u(\mathbf{x})}\equiv\ket{\boldsymbol{\Psi}(\mathbf{x})}\ \braket{u(\mathbf{x})|u(\mathbf{x})}_{\text{el}}=1\label{eq:EF definition}
\end{equation}
up to a \emph{gauge} choice, i.e. up to an arbitrary phase factor
$e^{i\varphi}$ ($e^{-i\varphi}$) which can be used to redefine $\ket{u(\mathbf{x})}$
($\psi(\mathbf{x})$) without affecting the total wavefunction \footnote{In Eq. \ref{eq:EF definition} the subscript $X$ (el) indicates that
integration is performed over nuclear (electronic) variables only.}. This introduces a proper (i.e., single-valued) electronic ``frame''
for each nuclear configuration. We assume that this can be done smoothly
in nuclear configuration space \textemdash{} hence, that the representation
of the Berry connection in the chosen frame, $A_{k}=i\braket{u|\partial_{k}u}$
for $k=1-N$, is well defined \textemdash{} at least in the regions
most relevant for the dynamics \footnote{In other words, we assume that we can select a normalized section
of the bundle that is smooth almost everywhere. Notice that there
is always the possibility of having ``essential'' discontinuities
that cannot be cured by any \emph{gauge} choice. This happens, for
instance, when $\ket{\Psi}$ represents an adiabatic state with a
CI, since in that case the electronic state is ill-defined right at
the CI seam. We assume $\ket{\boldsymbol{\Psi}(\mathbf{x})}=0$ there,
otherwise the nuclear derivatives $\partial_{j}\ket{\boldsymbol{\Psi}(\mathbf{x})}$'s
would be singular. }. At a wavefunction node $\ket{\boldsymbol{\Psi}(\mathbf{x})}\equiv0$,
the electronic state is arbitrary and we assume that it is selected
smoothly with its neighborhoods. Note that the total wavefunction
can be considered \emph{gauge} invariant, at least as long as the
molecule is isolated and does not couple to external fields (e.g.
an electromagnetic field). 

The evolution of the EF pair of functions can be obtained from either
the variational principle or projection operator techniques, as shown,
respectively, in Refs. \citep{Abedi2010,Abedi2012} or Refs. \citep{Martinazzo2022,Martinazzo2022a}.
It is not needed here, though, since the key quantities of interest
in this article require neither the evolution of the EF pair nor that
EF is actually performed. The key information is that an exact factorization
exists and that it introduces, at any time of interest, a proper electronic
frame in nuclear configuration space. Indeed, this implies a phase
quantization condition 
\begin{equation}
\sum_{k}\oint_{\gamma}p_{k}dx^{k}=\sum_{k}\oint_{\gamma}(\pi_{k}+\hbar A_{k})dx^{k}=2\pi\hbar\,n\label{eq:quantization condition}
\end{equation}
(with $n\in\mathbb{Z}$) which merely expresses the fact that the
EF nuclear wavefunction $\psi$ must be smooth around any loop $\gamma$
in nuclear configuration space. Here, with $\hat{p}_{k}=-i\hbar\partial_{k}$
denoting the Schr\"{o}dinger-representation canonical momentum, $p_{k}=\Re[(\hat{p}_{k}\psi)/\psi]\equiv\hbar\partial_{k}\text{arg \ensuremath{\psi}}$
is the $k^{\text{th}}$ derivative of the phase of the EF nuclear
wavefuntion and $\pi_{k}=p_{k}-\hbar A_{k}$ is the same component
of the mechanical momentum $\boldsymbol{\pi}$ appearing in the quantum
hydrodynamical description of the EF dynamics \citep{Martinazzo2023}.
Furthermore, $n$ is a topological value that describes the way the
wavefunction phase possibly winds around a singularity: it is non-zero
only if the singularity makes the domain multiply connected, otherwise
$\gamma$ can be shrunk to a single point and $n$ must vanish. Importantly,
the mechanical momentum and the quantization condition of Eq. \ref{eq:quantization condition}
are\emph{ gauge}-invariant, even though \emph{$p_{k}$ }and $A_{k}$
are not. In fact, $\boldsymbol{\pi}$ does not even require EF, it
can be obtained from the total $e-n$ wavefunction according to $\boldsymbol{\pi}\equiv\Re\braket{\boldsymbol{\Psi}(\mathbf{x})|\hat{\mathbf{p}}|\boldsymbol{\Psi}(\mathbf{x})}_{\text{el}}/n(\mathbf{x})$
\textendash{} where $\hat{\mathbf{p}}=(\hat{p}_{1},\hat{p}_{2},..\hat{p}_{N})$
and $n(\mathbf{x})=\braket{\boldsymbol{\Psi}(\mathbf{x})|\boldsymbol{\Psi}(\mathbf{x})}_{\text{el}}$
is the nuclear density.

The last observation, jointly with Eq. \ref{eq:quantization condition},
allows one to introduce an instantaneous geometric phase for arbitrary
paths, which reduces (modulo $2\pi$) to the holonomy of the EF vector
bundle when the path is closed (hence to the traditional Berry phase
when the state is adiabatic). Specifically, as shown in Ref. \citep{Martinazzo2023a},
one can decompose the ``total'' phase difference $\Theta_{\text{ba}}=\text{arg}\braket{\Psi(\mathbf{x}_{a})|\Psi(\mathbf{x}_{b})}$
between the endpoints $\mathbf{x}_{a}$ and $\mathbf{x}_{b}$ of a
curve $\gamma$ in nuclear configuration space as follows
\begin{equation}
\Theta_{ba}=\Gamma_{\text{n}}[\gamma]+\Gamma_{\text{el}}[\gamma]\label{eq:total phase difference}
\end{equation}
where 

\begin{equation}
\Gamma_{\text{el}}[\gamma]=\text{arg}\braket{u(\mathbf{x}_{a})|u(\mathbf{x}_{b})}+\sum_{k}\int_{\gamma}A_{k}dx^{k}\label{eq:open-path Berry phase}
\end{equation}
is the Pancharatnam phase accumulated by the electronic vector from
$\mathbf{x}_{a}$ to $\mathbf{x}_{b}$ along $\gamma$ \citep{Simon1983,Simon1993,Mukunda1993,Pati1995,Pati1998}
and 
\begin{equation}
\Gamma_{\text{n}}[\gamma]=\frac{1}{\hbar}\sum_{k}\int_{\gamma}\pi_{k}dx^{k}\label{eq:definition of QHD phase}
\end{equation}
is a nuclear phase. The decomposition is unique, since the above phases
are \emph{gauge} invariant, and it allows one to identify separate
electronic and nuclear contributions to the phase difference of the
total wavefunction. $\Gamma_{\text{el}}[\gamma]$ is indeed the phase
acquired by the electronic state vector in the \emph{parallel transport}
\emph{gauge} $\ket{\tilde{u}}$ defined by the condition $\sum_{k}\tilde{A}_{k}dx^{k}\equiv0$,
along the given curve \footnote{It is always possible to find such a \emph{gauge} for a given curve,
since $\ket{\tilde{u}}=e^{i\varphi}\ket{u}$ implies $i\sum_{k}\braket{\tilde{u}|\partial_{k}\tilde{u}}dx^{k}=-d\varphi+i\sum_{k}\braket{u|\partial_{k}u}dx^{k}$,
i.e. $\varphi=i\sum_{k}\int_{\text{\ensuremath{\gamma}}}A_{k}dx^{k}$.}. Likewise, $\Gamma_{\text{n}}[\gamma]$ is the phase accumulated
by the nuclear \emph{wavefunction} $\tilde{\psi}$ in the same \emph{gauge},
since $\sum_{k}\pi_{k}dx^{k}\equiv\sum_{k}\tilde{p}_{k}dx^{k}\equiv\hbar d(\text{arg}\tilde{\psi})$,
where tilde is used to denote quantities in the parallel transport
\emph{gauge}. 

Hence, for a given curve $\gamma$, $\Gamma_{\text{el}}[\gamma]$
and $\Gamma_{\text{n}}[\gamma]$ single out, from the infinitely many
\emph{gauges} made possible by the EF, a specific \emph{gauge} in
which monitoring the ``intrinsic'' phase changes of the two subsystems.
These correspond to \emph{wavefunction} phases in this and only this
gauge. This is also the \emph{gauge }where geometric phase effects
in the nuclear dynamics become manifest, since it is the most appropriate
for neglecting the vector potential in the EF effective Hamiltonian
$\tilde{H}^{\text{eff}}$ for the nuclear wavefunction \citep{Abedi2010,Abedi2012,Martinazzo2022,Martinazzo2022a},
at least when the focus is on the dynamics along the given (open)
path. When applying this approximation, i.e., when setting $A_{k}\equiv0$
in $\tilde{H}^{\text{eff}}$, the resulting Hamiltonian becomes equivalent
\emph{in form} to the Born-Oppenheimer Hamiltonian, provided one further
neglects the so-called diagonal corrections, i.e. disregard the effects
of the \emph{pseudo}-electric forces arising from the Fubini-Study
metric. This represents the point of closest contact with BO dynamics.
Notice however that the resulting effective Hamiltonian, freed of
geometric effects, contains yet time-dependent electronic states in
place of the stationary states that would appear in the BO limit. 

For a loop, if a geometric phase is present, the electronic frame
$\ket{\tilde{u}}$ is \emph{improper}, meaning that is not single-valued
anymore, and a smooth nuclear wavefunction satisfying the quantization
condition of Eq. \ref{eq:quantization condition} does not exist.
Rather, we have $-\Gamma_{\text{n}}^{O}[\gamma]=\Gamma_{\text{el}}^{O}[\gamma]=\text{arg}\braket{u_{a}|\tilde{u}_{a}}$
and $\text{arg}(\psi_{a}^{*}\tilde{\psi}_{a})=-\hbar\Gamma_{\text{el}}^{O}[\gamma]$
becomes the boundary condition appropriate for the nuclear wavefunction
in the parallel transport \emph{gauge} \footnote{To avoid confusion, the \emph{gauge} transformation needed to turn
a regular frame into a parallel transport \emph{gauge} is singular,
since it is not single-valued everywhere. Hence, the momentum field
and the quantization condition of Eq. \ref{eq:quantization condition}
do change  under this kind of transformations. }. This is also the case where departures from BO dynamics are larger,
since the latter cannot account for the multi-valuedness of the wavefunction.
Obviously, under such circumstances, a proper, smooth frame is more
convenient for describing the nuclear dynamics. With this choice $\Gamma_{\text{el}}^{O}[\gamma]$
becomes an integral property of the electronic frame and the nuclear
wavefunction satisfies the simpler condition of Eq. \ref{eq:quantization condition}. 

In practice, for a time-evolving wavepacket $\ket{\Psi_{t}}$ and
a path $\gamma_{t}$ \textemdash{} which possibly changes in time,
too \textemdash{} the total phase $\Theta_{ab}$ is directly obtainable
from the $e-n$ wavefunction amplitude at the endpoints of the path,
while $\Gamma_{n}[\gamma_{t}]$ is readily available from the instantaneous
momentum field, $\boldsymbol{\pi}(\mathbf{x},t)=\Re\braket{\boldsymbol{\Psi}_{t}(\mathbf{x})|\hat{\mathbf{p}}|\boldsymbol{\Psi}_{t}(\mathbf{x})}_{\text{el}}/n_{t}(\mathbf{x})$,
upon integrating the latter along the path at the given $t$, see
Eq. \ref{eq:definition of QHD phase}. Hence, the electronic phase
can be obtained from the above two quantities, without requiring any
explicit knowledge of a connection. The same momentum field $\boldsymbol{\pi}$
determines (through a mass tensor) the velocity field $\mathbf{v}(\mathbf{x},t)$
of the nuclear probability fluid, hence the QHD trajectories $\dot{\mathbf{x}}=\mathbf{v}(\mathbf{x},t)$
that dictate the evolution of any point $\mathbf{x}$ of a ``dynamical
object'' that moves in tandem with the fluid, for instance, our chosen
paths. We stress that these are Bohmian nuclear trajectories \citep{Holland1993}
in the EF-based representation of the dynamics, i.e. they describe
a Bohmian dynamics which accounts for the effects of both a quantum
potential and the non-conservative forces arising from the electron
dynamics. This settles the problem of obtaining geometric-phase related
quantities from any available exact solution of the dynamics.

\textbf{\emph{Model Jahn-Teller dynamics.}} We now focus on a 2-state
model that highlights the key features of a molecular problem involving
a CI. The nuclear system contains in general a number of degrees of
freedom described by $\mathbf{x}\in\mathcal{M}\cong\mathbb{R}^{N}$
and the electronic Hamiltonian takes the general form $H_{\text{el}}=A(\mathbf{x})\text{\ensuremath{\sigma_{0}}}+\mathbf{B}(\mathbf{x})\boldsymbol{{\sigma}}$,
where $A(\mathbf{x})$ is a scalar, $\mathbf{B}(\mathbf{x})\in\mathcal{N}\cong\mathbb{R}^{3}$
is an effective magnetic field, $\sigma_{0}=\mathbb{I}_{2}$ is the
2x2 unit matrix and $\boldsymbol{{\sigma}}=(\sigma_{x},\sigma_{y},\sigma_{z})$
is the vector of Pauli matrices. Here, we assume that a diabatic basis
\citep{Kouppel1984} $\{\ket{1},\ket{2}\}$ has been introduced and
represent the \emph{e-n} molecular wavefunction in spinor form. We
set $B_{z}\equiv0$ to mimic time-reversal invariance and make the
adiabatic Berry curvature vanishing everywhere except at the CI seam
(defined by the condition $\mathbf{B}(\mathbf{x})=0$). Under such
circumstances, the adiabatic Berry phase becomes topological (in fact,
it is the Longuet-Higgins phase \citep{Longuet-Higgins1958,Herzberg1963}),
namely $(\pm)\pi$ for any path in $\mathcal{N}$ space encircling
the CI once, when the system is in the ground (excited) state.  For
concreteness, we consider 2+1 nuclear degrees of freedom ($\mathcal{M}\cong\mathbb{R}^{3})$
mimicking a one-dimensional CI seam, with parameter values appropriate
for a molecular problem and a diagonal mass tensor, characterized
by the nuclear mass $M=1$ a.m.u.. Upon setting $A(\mathbf{x})=\frac{1}{2}M\omega_{x}^{2}x^{2}+\frac{1}{2}M\omega_{y}^{2}y^{2}$
and employing a linear vibronic coupling, $\mathbf{B}=\kappa_{x}x\mathbf{e}_{1}+\text{\ensuremath{\kappa}}_{y}y\mathbf{e}_{2}$,
the problem becomes effectively two-dimensional, in fact a standard,
linear $E\otimes e$ Jahn-Teller model, with the CI seam represented
by an isolated point, $\mathbf{x}=\mathbf{0}$. As for the parameters
of the Hamiltonian, we set $\omega_{x}=\omega_{y}=\omega=1000$ cm$^{-1}$
and $\kappa_{x}=\kappa_{y}=\kappa=0.1$ a.u..
\begin{figure*}
\begin{centering}
\includegraphics[width=1\textwidth]{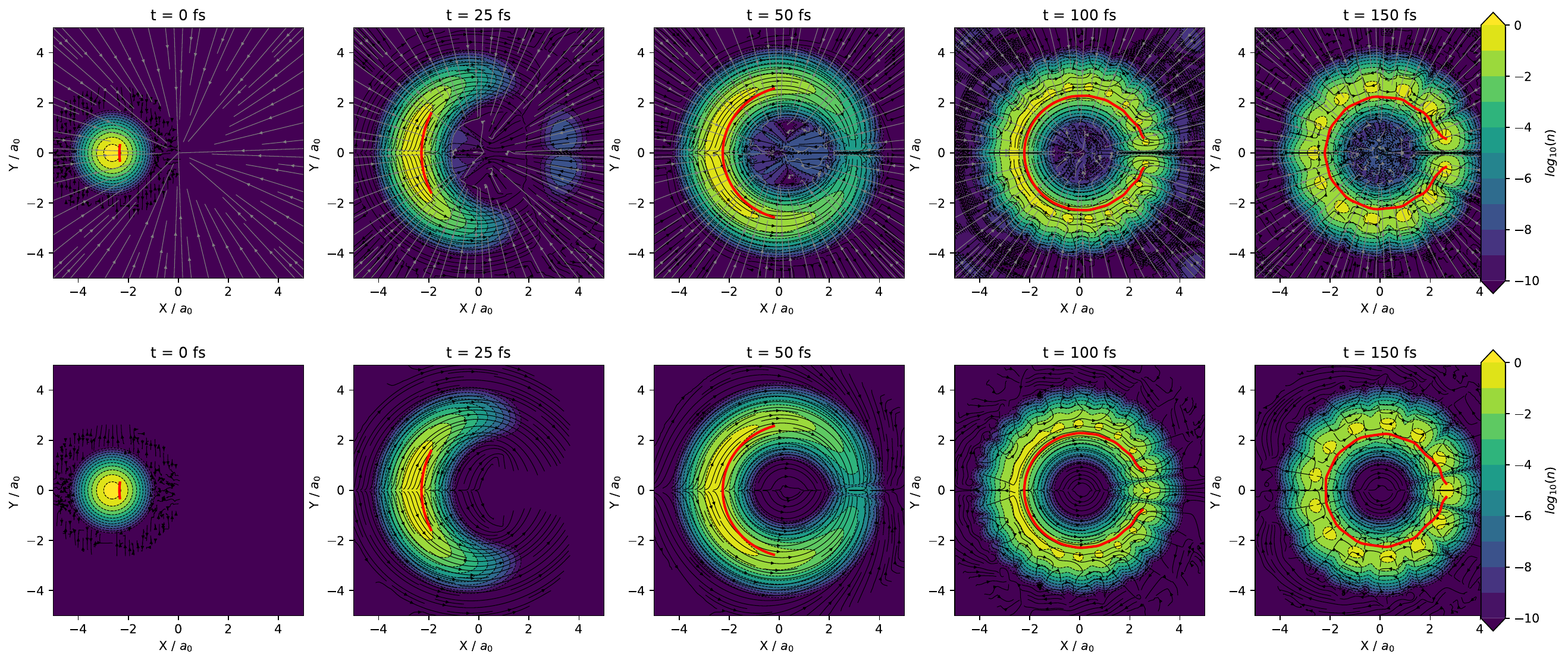}
\par\end{centering}
\caption{\label{fig:Snapshos (case a)}Evolution of an $e-n$ wavepacket encircling
a conical intersection point between two adiabatic electronic states,
located at the origin of the coordinate system. The top row shows
the evolution of the nuclear density (color coded in a log$_{10}$
scale according to the colorbar on the right) as follows from a numerically
exact solution of the time dependent Schr\"{o}dinger equation for
the 2-state model described in the main text (case (a)). The grey
lines with arrows represent the field lines of the polarization field
\textbf{s}, which gives information about the local electronic states
defined by the exact factorization of the wavefunction (see text for
details). The bottom row shows similarly the evolution of the nuclear
density in the Born-Oppenheimer approximation, when using the adiabatic
ground-state energy and the same initial nuclear state of the exact
dynamics. In each panel the red line represents the dynamical path
used for computing the phases of Eqs. \ref{eq:total phase difference}-\ref{eq:definition of QHD phase},
while the black lines with arrows are the field lines of the nuclear
momentum field $\boldsymbol{\pi}$. }
\end{figure*}
\begin{figure*}
\begin{centering}
\includegraphics[clip,width=1\textwidth]{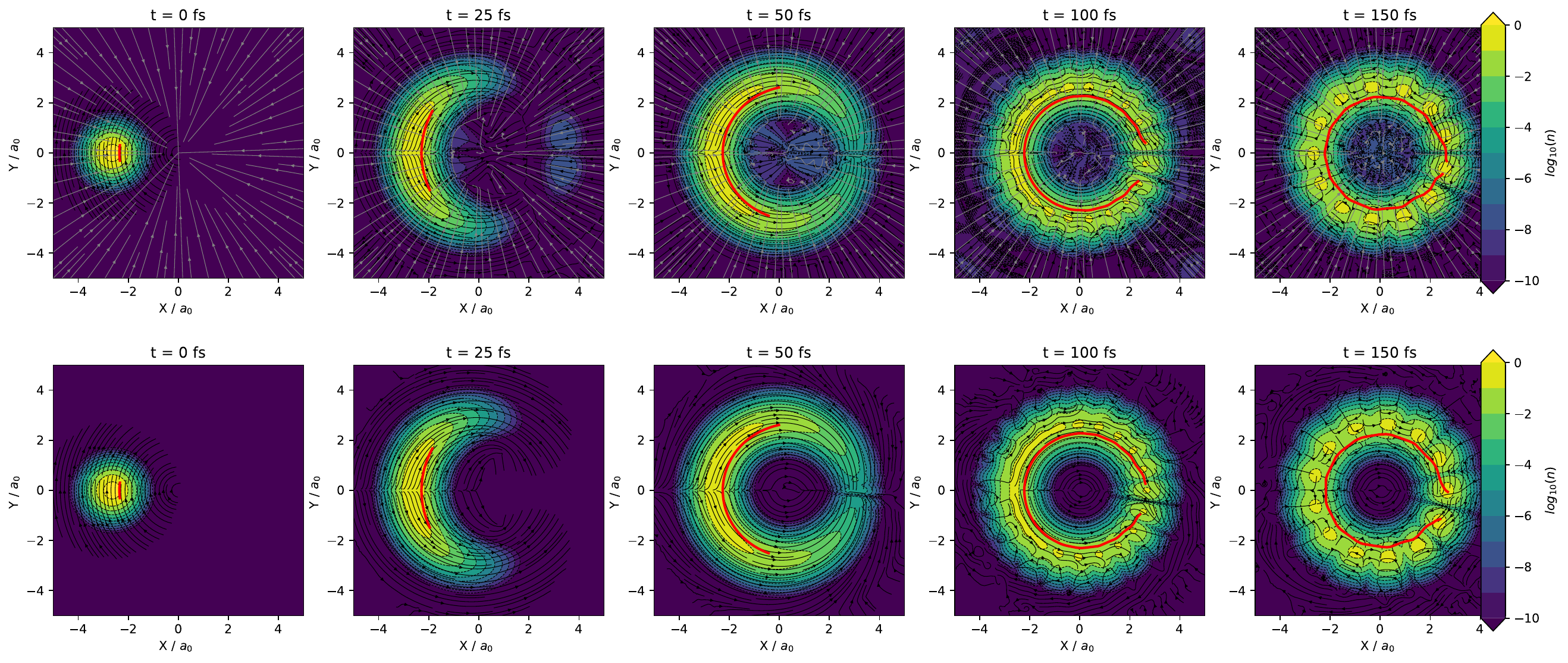}
\par\end{centering}
\caption{\label{fig:Snapshots (case b)}Same as in Fig. \ref{fig:Snapshos (case a)}
for case (b), where the chosen initial state has a drift term in the
clockwise direction (see main text for details). }
\end{figure*}

We solved the time-dependent Schr\"{o}dinger equation for this problem
using a standard Split-Operator algorithm in conjunction with Fast-Fourier-Transforms
to go back and forth between real- and momentum-space. The wavefunction
was represented on a fine grid (1024$\times$1024), which was centered
around the CI point and taken of length $20\,a_{0}$ along each direction.
A small time step of $\Delta t=0.1\,$a.u. was adopted to ensure a
good sampling of the geometric phases over time. The initial wavefunction
was obtained by combining a nuclear wavepacket $\psi_{0}(\mathbf{x})$
with a ground electronic state $\ket{u_{-}(\mathbf{x})}$, i.e. $\braket{\mathbf{x}|\Psi_{0}}_{X}=\psi_{0}(\mathbf{x})\ket{u_{-}(\mathbf{x})}$,
where $\psi_{0}(\mathbf{x})$ was a Gaussian centered at $x_{0}=-2\kappa/M\omega^{2}$
and $y_{0}=0$, with ground-state width along both $x$ and $y$ ($\Delta x=\Delta y=\sqrt{\hbar/2m\omega})$
and zero nominal momentum along both directions. Two slightly different
initial states were considered, both using the same $\psi_{0}(\mathbf{x})$
just described and differing only for the \emph{gauge} choice of $\ket{u_{-}(\mathbf{x})}$
(see below). They both give rise to a dynamics which is very close
to adiabatic, such that a comparison with Born-Oppenheimer dynamics
is meaningful. Therefore, in parallel, we further performed calculations
in the BO approximation, employing the same set-up described above.
To this end we set $\mathbf{B}(\mathbf{x})=B(\mathbf{x})\sigma_{z}$
(where $B$ is the magnitude of $\mathbf{B}$) and used a fictitious
electronic state, namely $\boldsymbol{\chi}(\mathbf{x})\equiv[0,1]$
in the chosen diabatic basis. As for the initial state wavefunction
$\phi_{0}(\mathbf{x})$ appropriate for this problem we made sure
that it represented the \emph{same} (nuclear) state used in the exact
problem, as described in the following. 

A first initial state, henceforth case (a), was obtained with the
\emph{gauge} choice $\ket{u_{-}(\mathbf{x})}=(e^{-i\phi/2}\ket{1}-e^{i\phi/2}\ket{2})/\sqrt{2}$
where $\phi$ is the azimuthal angle of the position vector $\mathbf{x}$.
Even though this is an improper electronic frame presenting a discontinuity
on the positive $x$ axis this is irrelevant for our purposes, since
$\psi_{0}(\mathbf{x})$ localizes on the negative $x$ axis. In fact,
this is a parallel transport frame for any open path not crossing
the positive $x$ axis, since the corresponding Berry vector potential
vanishes everywhere in $\mathbb{R}^{2}$ except on that semi-axis,
where it is singular (in $\mathbb{R}^{3}$ this would be the direction
of the so-called Dirac line). The corresponding BO counterpart of
this initial state was then set to $\phi_{0}(\mathbf{x})\equiv\psi_{0}(\mathbf{x})$.
This choice ensures ensures that the two dynamics share the same density
$n(\mathbf{x})$ and the same momentum field $\boldsymbol{\pi}(\text{x})$
at the initial time \footnote{Strictly speaking this is true \emph{almost} everywhere and, in particular,
in the regions of nuclear configuration space where it is most relevant,
i.e. where $n(\mathbf{x})$ is sizable.}. Notice that $\boldsymbol{\pi}(\text{x})\equiv\mathbf{0}$ at $t=0$
(where $\boldsymbol{\pi}$ is well-defined), since $\psi_{0}(\mathbf{x})$
is real and there is no contribution to the mechanical momentum coming
from the Berry connection. This defines a state that remains symmetric
(even) with respect to mirror reflections about the $x$ axis (see
Fig. \ref{fig:Snapshos (case a)}). 

In a second initial state, henceforth case (b), the \emph{gauge} was
fixed with the proper choice $\ket{u_{-}(\mathbf{x})}=(-\ket{1}+e^{i\phi}\ket{2})/\sqrt{2}$,
which is a special case of a standard parametrization of the Bloch
sphere, often used to describe the ground state of a spin in a magnetic
field. This parametrization reads as $\ket{u_{-}}=\sin(\theta/2)\ket{1}-e^{i\phi}\cos(\theta/2)\ket{2}$,
for the field along the unit vector $\mathbf{b}=\cos\phi\sin\theta\mathbf{e}_{1}+\sin\phi\sin\theta\mathbf{e}_{2}+\cos\theta\mathbf{e}_{3}$,
where $\theta$ and $\phi$ are the usual latitude and longitude angles
of the spherical coordinates in $\mathbb{R}^{3}$ \footnote{The more common parametrization is for the $+$ (i.e. excited) state,
where the spin is directed along the field axis. It is related to
the one given here by a reversal of $\mathbf{b}$, i.e. $\theta\rightarrow\pi-\theta$
and $\phi\rightarrow\phi+\pi$. }. It is regular everywhere on the Bloch sphere except for $\theta=0$,
hence the Dirac line of the corresponding Berry vector potential lies
along the positive $z$ axis and it is harmless for our problem (where
$\theta\equiv\pi/2$). This comes at a price of introducing a non-vanishing
vector-potential contribution to the initial momentum $\boldsymbol{\pi}$,
a clockwise vortex structure centered at the CI point, that causes
a drift of the system in the same sense \footnote{A simple calculation shows that $\boldsymbol{\pi}=-\hbar/2r\,\mathbf{e}_{\phi}$,
where $r=|\mathbf{x}|$ and $\mathbf{e}_{\phi}$ is the unit vector
along the $\phi$ coordinate.} (Fig. \ref{fig:Snapshots (case b)}). In this case, $\ket{\tilde{u}_{-}}=e^{-i\phi/2}\ket{u_{-}}$
would be needed to define the parallel transport \emph{gauge}, hence
$\phi_{0}(\mathbf{x})=\tilde{\psi}_{0}(\mathbf{x})=e^{i\phi/2}\psi_{0}(\mathbf{x})$
is our corresponding choice of the initial state for the BO dynamics.
Again, this ensures that not only the same nuclear density $n(\mathbf{x})=|\psi_{0}(\mathbf{x})|^{2}$
but also the same momentum field $\boldsymbol{\pi}(\text{x})$ are
defined at initial time for the two dynamics. 

The resulting dynamics is very similar in the two cases: the wavepacket,
initially located on one side of the CI point, spreads along the valley
of the \textquotedblleft Mexican hat\textquotedblright{} potential
and its trailing edges meet each other and start interfering at time
$t\approx50\,fs$, after which the wavepacket covers more or less
uniformly the valley, with a time-varying interference pattern (Figs.
\ref{fig:Snapshos (case a)} and \ref{fig:Snapshots (case b)}). The
dynamics is adiabatic to a large extent, as a visual inspection of
the field lines of the polarization field $\mathbf{s}(\mathbf{x})$
suggests and a more detailed analysis of the populations confirms
(not reported). The vector field $\mathbf{s}(\mathbf{x})$ characterizes,
in a \emph{gauge} invariant way, the local electronic states defined
by the EF of the total wavefunction, namely \emph{via} $\rho_{\text{el}}(\mathbf{x})=(\text{\ensuremath{\mathbb{I}}}_{2}+\mathbf{s}(\mathbf{x})\boldsymbol{\sigma})/2$,
where $\rho_{\text{el}}(\mathbf{x})$ is the matrix representation
(in the chosen diabatic basis) of the conditional density operator
$\hat{\rho}_{\text{el}}(\mathbf{x})=\ket{u(\mathbf{x})}\bra{u(\mathbf{x})}$,
$\text{\ensuremath{\mathbb{I}}}_{2}$ is the unit 2$\times$2 matrix
and $\boldsymbol{\sigma}$ is the vector of Pauli matrices (for an
adiabatic ground / excited state the polarization field reads, in
our set-up, as $\mathbf{s}_{\mp}(\mathbf{x})=\mp\mathbf{x}/\lVert\mathbf{x}\rVert$).
The interference patterns in Figs. \ref{fig:Snapshos (case a)} and
\ref{fig:Snapshots (case b)} depend on the details of the initial
state, as the evolution of a dynamical path tied to the nuclear probability
density makes evident (red line). Importantly, the pattern is reversed
compared to the BO dynamics, a clear signature of geometric (topological)
phase effects. The subtle differences between case (a) and case (b)
(at the same level of description) can be revealed (by inspection)
only in the path dynamics, and follow from the behavior of the phases
introduced in the previous Section, as we now show.
\begin{figure*}
\begin{centering}
\includegraphics[width=0.9\textwidth]{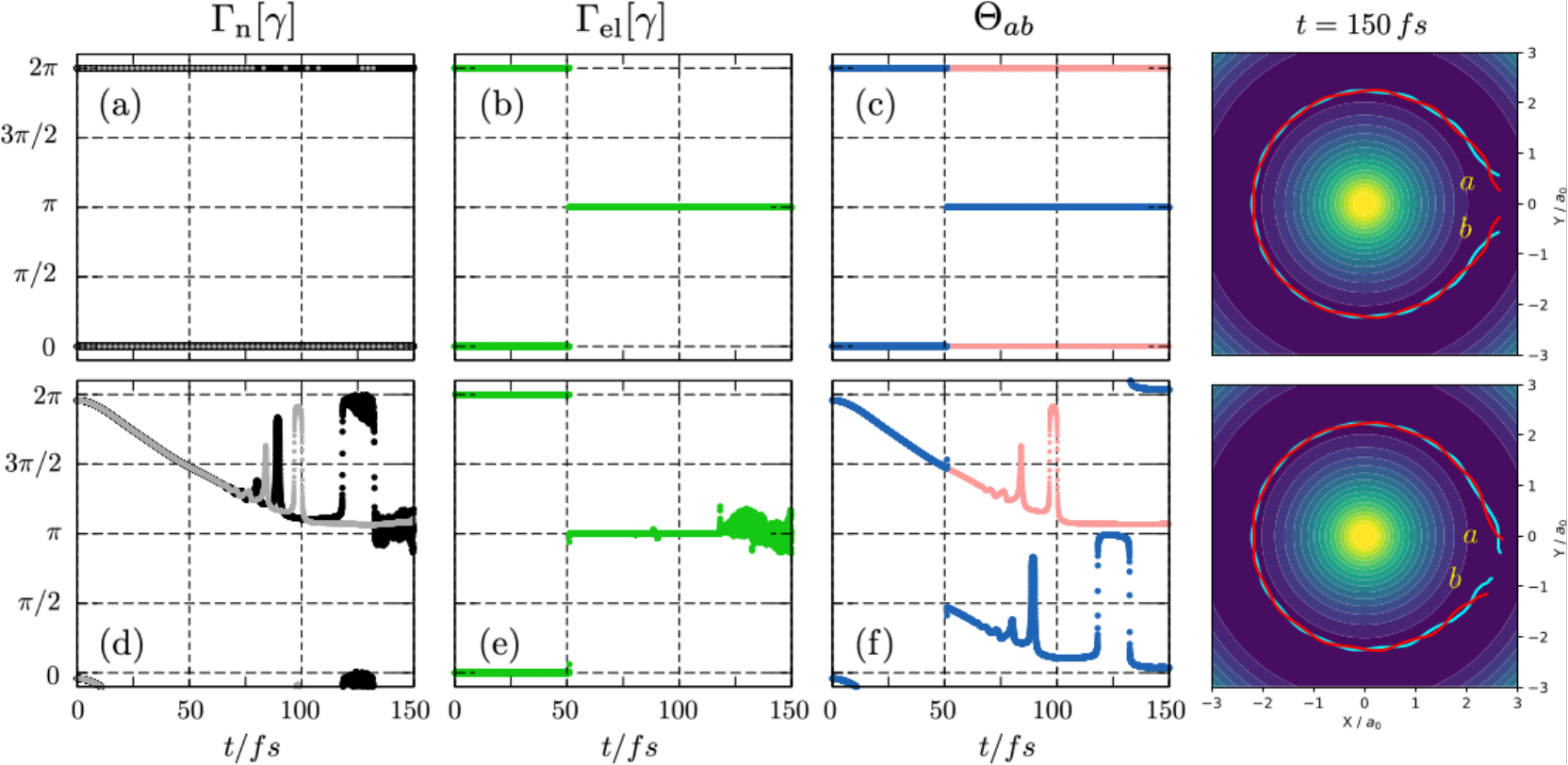}
\par\end{centering}
\caption{\label{fig:Phase-differences}Phase differences between the endpoints
$a$ and $b$ of the dynamical paths shown in Figs. \ref{fig:Snapshos (case a)}-
\ref{fig:Snapshots (case b)}. Panels (a-c) for Fig. \ref{fig:Snapshos (case a)}
and panels (d-f) for Fig. \ref{fig:Snapshots (case b)}. Panels (a)
and (d) display the nuclear phase $\Gamma_{\text{n}}[\gamma]$ (black),
(b,e) the electronic phase $\Gamma_{\text{el}}[\gamma]$ (green symbols)
and (c,f) the total phase difference $\Theta_{ab}$ (light blue).
Also shown are the results obtained from the Born-Oppenheimer dynamics,
grey symbols in (a,d) and red symbols in (c,f). Data are shown with
their $\pm2\pi$ images for displaying purposes. The two rightmost
panels show the paths at $t=150\,fs$, for the exact (cyan) and for
the Born-Oppenheimer (red) dynamics, superimposed to a color map of
the adiabatic, ground-state potential energy surface and with $a$,
$b$ labeling the endpoints of the curves.}
\end{figure*}

Fig. \ref{fig:Phase-differences} shows the nuclear ($\Gamma_{\text{n}}[\gamma]$),
electronic ($\Gamma_{\text{el}}[\gamma]$) and total ($\Theta_{ab}$)
phase difference between the end points of the probe path that moves
in tandem with the nuclear probability fluid and that was started
as an arc centered at the initial wavepacket position (leftmost panels
in Figs. \ref{fig:Snapshos (case a)} and \ref{fig:Snapshots (case b)}).
The behavior of the electronic phase \textendash{} only relevant for
the exact dynamics \textendash{} is essentially the same, irrespective
of the initial state, for our quasi-adiabatic dynamics around the
CI point. The phase remains constant up to about $50\,fs$, when the
path becomes approximately a semicircle and the phase changes abruptly
to $\pi$ (mod$2\pi$). This is also the limiting value that the phase
takes when artificially closing the path after the transition point
(i.e., when closing the path encompasses the CI point), and $\Gamma_{\text{el}}[\gamma]$
is seen to keep this value for $t\geq50\,fs$. This distinctive behavior
of $\Gamma_{\text{el}}[\gamma]$ is a feature of the Pancharatnam
phase of Eq. \ref{eq:open-path Berry phase} that is easily seen to
be operative already on the Bloch sphere, when parallel transporting
vectors along great circles. For instance, when parallel transporting
a ground-state vector $\ket{u_{a}}$ along the $\phi_{0}$ meridian
line $[0,1]\ni s\rightarrow(s\pi,\phi\equiv\phi_{0})$, up to the
south pole, and back to the north pole along the antimeridian line
$[0,1]\ni s\rightarrow(\pi(1-s),\phi\equiv\phi_{0}+\pi)$ to give
$\ket{\tilde{u}_{a}}=-\ket{u_{a}}$. Indeed, using the above mentioned
parameterization of the Bloch sphere \textendash{} i.e., $\ket{u_{-}}=\sin(\theta/2)\ket{1}-e^{i\phi}\cos(\theta/2)\ket{2}$
\textendash{} the line integral does not contribute to the phase since
$\mathbf{A}$ lies on latitude lines while, on the other hand, $\text{arg}\braket{u_{-}(0,\phi_{0})|u_{-}(\theta,\phi)}$
undergoes a sudden $\pi$ jump when traversing the south pole. Changing
the \emph{gauge} makes the calculation a bit longer but does not affect
the final result, since $\Gamma_{\text{el}}[\gamma]$ is \emph{gauge}
invariant. 

Now, in the previous example the south pole is the point where the
transported vector $\ket{\tilde{u}_{b}}$ is orthogonal to $\ket{u_{a}}$
and the phase $\arg\braket{u_{a}|\tilde{u}_{b}}$ is ill defined.
This is the way the non-trivial Berry phase is generated when traveling
along the great circle. In other words, the $-1$ (topological) phase
factor does not build up continuously along the path, rather it is
suddenly acquired when the state vector becomes orthogonal to the
initial one. This is precisely what happens with the EF local electronic
states attached to the endpoints of our dynamical path, when the latter
is stretched for approximately half of a circle length along the valley
of the ground-state potential, since the dynamics is essentially adiabatic
and $\Gamma_{\text{el}}[\gamma]$ tracks the phase changes in the
parallel transport \emph{gauge}. 

The behavior of the nuclear phase $\Gamma_{n}[\gamma]$, on the other
hand, does depend on the initial state, being intimately related to
the ensuing nuclear dynamics which is slightly different in the two
cases, because of the absence (case (a)) or presence (case (b)) of
a non-vanishing momentum field $\boldsymbol{\pi}$ at initial time.
In case (a) the line integral of Eq. \ref{eq:definition of QHD phase}
vanishes by symmetry, whereas in case (b) the endpoints of the path
travel at different velocities and cancellation of terms cannot occur.
The behaviour is however essentially the same as in the Born-Oppenheimer
dynamics, at least as long as the path endpoints \textemdash which,
we recall, are closely related to the trailing edges of the wavepacket
\textemdash{} are sufficiently far apart. This is due to the fact,
mentioned above, that the phase takes the meaning of nuclear wavefunction
phase in the parallel transport \emph{gauge}, where the effective
Hamiltonian $\tilde{H}^{\text{eff}}$ describing the evolution of
the marginal probability amplitude in the EF approach looks very similar
to a BO Hamiltonian. This implies that the two dynamics are (locally)
very similar to each other. Notice that for the problem discussed
here the parallel transport condition (in the adiabatic limit) is
topological, too, meaning that $\mathbf{A}=\mathbf{0}$ is a condition
that defines a ``global'' parallel transport \emph{gauge, }that
does not depend on the chosen path \footnote{The condition $\nabla\times\mathbf{A}=\mathbf{0}$ allows one to write
$\mathbf{A}=\nabla\varphi$ in any simply connected subdomain, where
$\varphi$ is a scalar. Hence, for any path \emph{not} encircling
the CI, $\ket{u}\rightarrow e^{i\varphi}\ket{u}$ defines the parallel
transport \emph{gauge}, $\mathbf{A}\rightarrow-\nabla\varphi+\mathbf{A}=\mathbf{0}$.}. 

In fact, the exact dynamics starts departing from the BO dynamics
only when the trailing edges of the wavepacket come close to each
other and interfere, as dictated by the \emph{total} phase difference
$\Theta_{ab}$ (Fig. \ref{fig:Phase-differences} (c,f)). For the
BO dynamics $\Theta_{ab}\equiv\Gamma_{n}[\gamma]$ (in Fig. \ref{fig:Phase-differences}
to within the numerical accuracy of the computation of the line integral
of Eq. \ref{eq:definition of QHD phase}) while for the exact dynamics
$\Theta_{ab}\approx\Gamma_{n}[\gamma]+\pi$. Hence, in case (a) (see
Fig. \ref{fig:Phase-differences} (c)) the wavapacket develops a node
in the exact dynamics which is absent in the BO dynamics. This is
manifest in the different behavior of the path endpoints, which is
summarized in the top right panel of Fig. \ref{fig:Phase-differences}.
On the other hand, in case (b) (see Fig. \ref{fig:Phase-differences}
(f)) it is the BO dynamics that presents a node, since in the exact
dynamics the edges of the wavepacket can be smoothly merged due to
$\Theta_{ab}\approx0$ (mod$2\pi$). This is clearly seen in the maps
shown in the rightmost panels of Figs. \ref{fig:Snapshos (case a)}
and \ref{fig:Snapshots (case b)}, that display the nuclear density
(in a log$_{10}$ scale) at long time.

We stress that the phases of Fig. \ref{fig:Phase-differences} are
\emph{open}-path phases, for which the parallel transport \emph{gauge}
remains smooth all over the path and $\Gamma_{\text{n}}[\gamma]$,
$\Gamma_{\text{el}}[\gamma]$ can be associated to single-valued nuclear
and electronic wavefunctions. Whether we artificially close the paths
by bringing the endpoints $a$ and $b$ closer together, the \emph{limiting}
total phase difference $\Theta_{ab}$ depends on the possible presence
of a nodal line crossing the path, $\Theta_{ab}\rightarrow0$ without
nodes and $\Theta_{ab}\rightarrow\pi$ for a node \footnote{Notice that a node $\ket{\Psi(\mathbf{x})}=0$ makes the phase $\Theta_{ab}=\text{arg}\braket{\Psi(\mathbf{x}_{a})|\Psi(\mathbf{x})}$
no longer continuous.}. This ``$\pi$ contribution'' appearing in $\Theta_{ab}$ when
a node is present parallels the $\delta-$like term that arises in
the momentum field $\boldsymbol{\pi}$ along a \emph{closed}-path
that crosses a nodal line. For instance, if $\psi(x,y)=0$ in some
interval of the positive $x$ axis, then $\psi(x,y)\approx a(x)y$
holds around that axis (for $a(x)=(\partial_{y}\psi)(x,0)$) and $\pi_{y}=\hbar\Im(1/y)=\pm\hbar\pi\delta(y)$,
when the singularity is circumvented by deforming the path in the
complex $y$ plane. This means that, when a node develops between
the endpoints of the curve, a $\pi$ contribution would emerge from
the very act of closing the path, i.e., when replacing $\Gamma_{\text{n}}[\gamma]$
with $\Gamma_{\text{n}}^{O}[\gamma]$. This would occur for the exact
dynamics in panel (a) of Fig. \ref{fig:Phase-differences} and for
the BO dynamics in panel (d) of the same figure, and would formally
enforce the condition $\Theta_{ab}=0$, irrespective of the presence
or absence of a nodal line. The implicit assumption is that the EF
electronic state at the node (that, we recall, is arbitrary) is chosen
smoothly, such that $\lim_{b\rightarrow a}\Gamma_{\text{el}}[\gamma]=\Gamma_{\text{el}}^{O}[\gamma]$. 

Overall, the above results confirm elementary arguments about the
topological contributions of the electronics to the total phase difference
between the trailing edges of a wavepacket encircling the CI. The
choice of the initial state, though, is seen to play a distinctive
role in determining whether a perfect destructive interference can
or cannot occur between the edges. Noteworthy is that in case (b),
an ``attempt'' to form a node is evident at $t\approx125\,fs$ (Fig.
\ref{fig:Phase-differences}, panels (d) and (f)), but this fails
since the broken symmetry of the initial state prevents perfect cancellation
of contributions. 

Finally, we remark that it is the existence of situations like cases
(b) that makes inclusion of geometric phases not straightforward in
typical mixed quantum-classical methods \citep{Nelson2020a,Shu2023}
where electrons are treated quantally and nuclei classically. 

\textbf{\emph{Conclusions.}} We have investigated molecular geometric
phase effects from an \emph{exact} quantum dynamical perspective,
by defining \emph{gauge} invariant nuclear and electronic phase differences,
which are valid for arbitrary electronic states and for arbitrary
paths in nuclear configuration space. We have shown that these phases
may reveal key information about the wavefunction dynamics and the
effects of a geometric phase, thereby providing a firm basis for the
existence of geometric phase effects in molecules beyond the adiabatic
approximation, and support for the elementary arguments often invoked
in their explanation. 

%
\pagebreak
\end{document}